# Compact and explicit physical model for lateral metal-oxide-semiconductor field-effect transistor with nanoelectromechanical system based resonant gate

L. Duraffourg, E. Colinet, S. Hentz, E. Ollier, P. Andreucci, B. Reig, P. Robert

CEA-LETI – MINATEC – 17 rue des Martyrs 38054 GRENOBLE Cedex 09 – France

We propose a simple analytical model of a metal-oxide-semiconductor field-effect transistor (MOSFET) with a lateral resonant gate based on the coupled electromechanical equations, which are self-consistently solved in time. All charge densities according to the mechanical oscillations are evaluated. The only input parameters are the physical characteristics of the device. No extra mathematical parameters are used to fit the experimental results. Theoretical results are well in agreement with experimental data in static and dynamic operation. Our model is comprehensive and may be suitable for any electromechanical device based on the field effect transduction.

Field effect transduction is certainly among the older techniques used in micro electromechanical systems (MEMS) [1]. However, it had rapidly been supplanted by capacitive techniques that use simple MEMS technologies. It is commonly admitted that capacitive detection with a low noise amplifier exhibits ultra low noise. Today, these noise levels are sufficiently low to reach the resolution needed for typical MEMS sensors such as inertial sensors, or pressure sensors [2]. Now, NEMS (Nano-Electro-Mechanical-Systems) devices are actively investigated because of their physical properties resulting from ultra miniature size elements [3]. Furthermore, they offer the opportunity to integrate mechanical structures and CMOS devices on the same die. NEMS advantages include ultra low power consumption, potential high resonant frequency, and high sensitivity to applied force, external damping or additional mass [4] [5].

Here, we study a detection technique consisting in a nano structure oscillating along the channel of the MOSFET. The lateral resonant gate MOSFET (LRG-MOSFET) and the CMOS circuit can be fabricated on the same die through thin SOI technology and SON (Silicon on Nothing) technology [6]. The same thin layer of single-crystal silicon is used for both mechanical structures and advanced ICs. In the resonant gate MOSFET, the gate moved along the channel width [7]. The drain current was then easily calculated by considering N basic transistors placed in parallel along its width. This paper is devoted to the modelling of LRG-MOSFET. As the suspended gate moves along the channel length, the developed approach in [7] is no longer usable and another electromechanical model has to be developed. A self-consistent model computing both mechanical gate deflection and surface potential onto the channel is therefore presented. The electrical model is based on an explicit formulation of the surface potential to determine all charge densities in the MOSFET. The fringe effect and the mode shape of the beam are included in the computation of the electrostatic forces. Theoretical results are finally compared with raw data measured on a device.

The geometry of the component and the notations are showed in the Figure 1

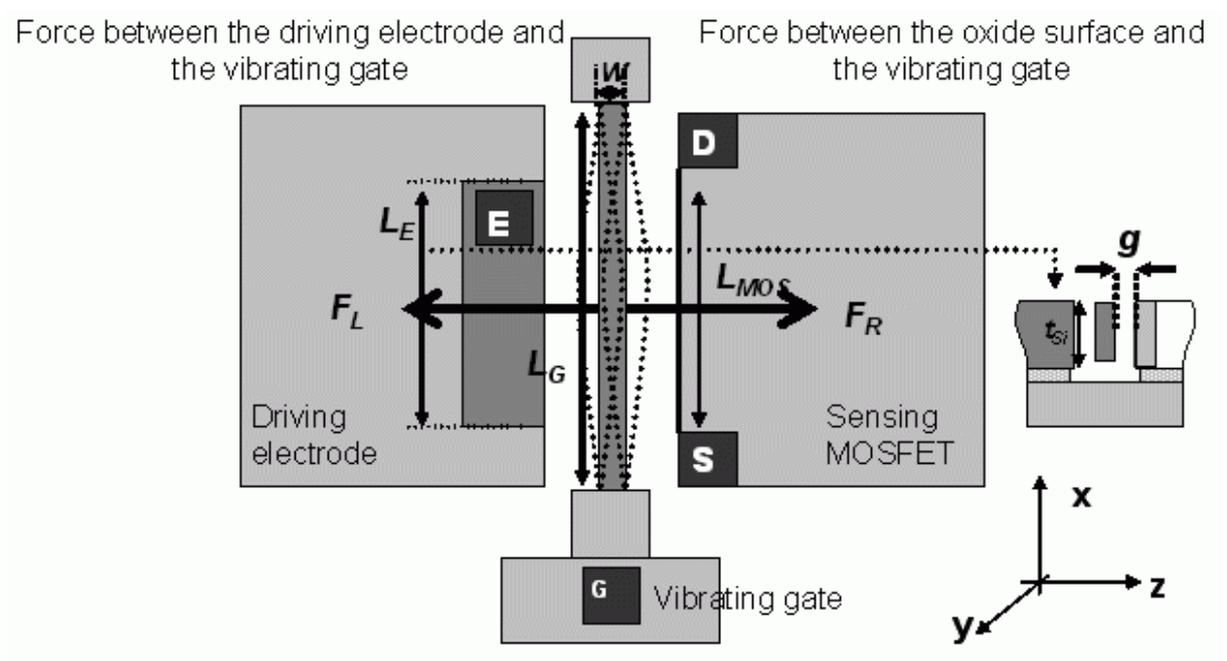

Figure . Second Newton equation is applied in the Galilean referential frame of the substrate to the ensemble beam+MOSFET. The beam is assumed to follow Euler-Bernouilli equation [8]. The equation is reduced to its normalized lumped expansion on the first mode through the Galerkin method, which consists in writing the solution over an orthogonal modal base $\{\chi_n(x)\}_{n=1,\infty}$:

$$\ddot{z}_1 + \omega_1^2 z_1 + \alpha_3 z_1^3 + \frac{\omega_1}{Q}\dot{z}_1 = (F_R - F_L) \tag{1}$$

The cubic term $\alpha_3$ is the Duffing effect, Q is the quality factor, $\omega_1$ is the free resonance frequency of the first mode. $F_R$ and $F_L$ correspond to the electrostatic forces acting on each side of the beam. Labels L and R are for left and right sides (see

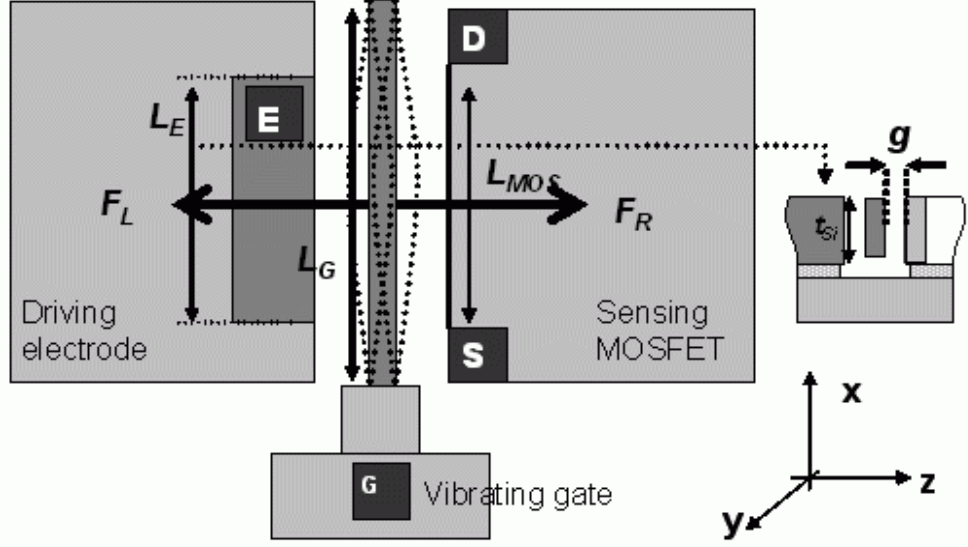

Figure ). $\{\chi_n(x)\}_{n=1,\infty}$ follows the normalisation condition:

$$\int_0^{L_G} \rho S \chi_n(x) \chi_m(x) dx = \delta_{nm} \quad (2)$$

The electrostatic force densities are projected on $\chi_1(x)$:

$$F_L = \frac{V_L^2}{2} \int_0^{L_E} \frac{\partial C_L}{\partial z_1} \chi_1(x) dx \text{ with } V_L = V_g - V_E \quad (3)$$

$$F_R = \frac{1}{2} \int_0^{L_{MOS}} V_R^2 \frac{\partial C_R}{\partial z_1} \chi_1(x) dx \text{ with } V_R = V_g - V_{oxide}(z(x)) \quad (4)$$

$V_g$, $V_E$, $V_{oxide}$ are respectively the gate voltage, the electrode voltage and the voltage between the surface of the oxide and the ground. $C_L$ and $C_R$ are the air gap capacitances per length unit:

$$C_{L,R}(x) = \frac{C_n \varepsilon_0 t_{Si}}{(g \pm z(x))} \quad (5)$$

$\varepsilon_0$ is the vacuum permittivity. $C_n$ is the fringe effect factor [9].

$F_R$ is a bit more complex than $F_L$ because the potential difference depends explicitly on $z(x)$ through $V_{oxide}$, which depends on the capacitive bridge $V_g \Big/ 1+\dfrac{C_{MOS}}{C_R}$. The classical MOSFET equations for each z-displacement are used to define $C_{MOS}$.

First, the surface potential $\varphi_s(x)$ is computed with an explicit model, which is obtained by asymptotic simplifications of the implicit equation for each operating regime [10]:

$$\left(\dfrac{V_g - V_{FB} - \varphi_s}{\gamma}\right)^2 = \varphi_s + \dfrac{k_B T}{q}\left[\exp\left(\dfrac{-q\varphi_s}{k_B T}\right) - 1\right] + \dfrac{k_B T}{q}\left[\exp\left(\dfrac{q\varphi_s}{k_B T}\right) - 1\right]\exp\left(\dfrac{-q(V + 2\phi_{Fi})}{k_B T}\right) \quad (6)$$

$$\gamma = \sqrt{\dfrac{2q\varepsilon_{Si}\varepsilon_0 N_a}{C_i^2}}, \text{ where } C_i = \left(\dfrac{t_{ox}}{\varepsilon_{ox}} + \dfrac{1}{C_R}\right)^{-1}$$

$C_i$ is the dielectric capacitance due to the oxide layer and the variable air gap. $V_{FB} = \phi_{ms} - \dfrac{Q_i}{C_i} - \dfrac{Q_{ss}}{C_i}(\varphi_s - \phi_{Fi})$ is the flat band voltage. $Q_i$ and $Q_{ss}$ are respectively the charge density in the gate oxide and the charge density at the interface oxide / silicon. $V(x)$ is the potential in the channel depending on $V_{ds}$. $q$, $N_a$, $\varepsilon_{Si}$, $k_B$, $T$ and $\phi_{Fi}$ are respectively the electron charge, the doping level of the channel, the relative dielectric permittivity, the Boltzmann constant, the temperature and the Fermi potential.

Once $\varphi_s$ known, the charge densities in the MOSFET and the MOS capacitance can be evaluated to determine oxide voltage and drain current:

$$Q_{dep}(x) = -\gamma C_i \sqrt{\varphi_S(x) + \dfrac{k_B T}{q}\left[\exp\left(\dfrac{-q\varphi_s(x)}{k_B T}\right) - 1\right]} \dfrac{(V_g - V_{FB})}{|V_g - V_{FB}|} \quad (7)$$

$$C_{sc}(x) = \dfrac{\partial Q_{sc}(x)}{\partial \varphi_s}; C_{MOS}(x) = \left(\dfrac{1}{C_{sc}} + \dfrac{t_{ox}}{\varepsilon_{ox}}\right)^{-1} \quad (8)$$

Where, $Q_{dep}(x)$, $Q_{inv}(x)$ and $Q_{sc}(x)$ are respectively the depletion charge density, the inversion charge density and the charge density in the gate. The drain current is the sum of the inversion charge density along the channel.

$$Id = -\mu_{eff} \frac{W}{L} \int_{V_s}^{V_d} Q_{inv}(V) dV \qquad (9)$$

$\mu_{eff}$ is the effective mobility of electron (N-MOSFET). For each operating point ($V_g$, $V_{ds}$), $\mu_{eff}$ is approximated by the Mathiessen rule considering the phonon interaction, the Coulomb interaction and the surface scattering [11].

Finally, the model depends on the geometry and a set of physical parameters ($\mu_{eff}$, $Q_i$, $Q_{ss}$ and $N_a$). The equations (1) to (8) have to be solved self consistently through a robust iterative algorithm based on the Runge-Kutta formulation.

To validate our theoretical model, we compare the simulation results with measurements reported in [12]. Devices were fabricated with a technology based on 200mm tools with a SON (Silicon on Nothing) approach [7]. This technology takes advantage of the single-crystal silicon for the NEMS structure without using SOI wafers. This technology is compatible with a front-end CMOS process. The Figure 2 shows a SEM micrograph of a typical device.

Let us consider a device (Figure 2) on which static measurement were carried out. The measured characteristics are $L_G$=16.1 µm, w=490 nm, g=107 nm, $L_{MOS}$= 9 µm, $W_{MOS}$=400 nm, $t_{ox}$=2 nm, and $N_a$= $5.10^{15}$ at/cm$^3$ [12]. Figure 3 a) and b) give respectively the static characteristics $I_d(V_g)$ and $g_m(V_g)$ with $V_E$=$V_S$=0. The theoretical results are superimposed in each case. Notice that the bulk voltage (body potential) remains floating. To fit the model with the experimental results, the doping level $N_a$, and the charge density in the gate oxide $Q_i$ are respectively fixed at $4.5.10^{15}$ at/cm$^3$ and $5.10^{10}$ cm$^{-2}$ since the charge density $Q_{os}$ at the interface oxide / silicon is tuned. This parameter depends on the technology (etching step and oxide deposition) as well as the voltages. In the Figure 3 a), theory and measurement are well in agreement. Threshold voltages are quite similar around 2 V. Below the threshold, in the accumulation regime, the experimental device exhibits a quite strong leakage current from 0.1 µm to 1.3µA. This current variation depends linearly on $V_{ds}$ at $V_g$=0V. It may be attributed to a photolithography misalignment when protecting channel from phosphorous implants generating a 2.1 MΩ short-circuit resistor. This process error can not be reproduced by the model and the resistance was added by hand in parallel of the transistor. For $V_{ds}$=1.55V and 2.75V, the charge density $Q_{ss}$ is respectively fixed at $10^{12}$ cm$^{-2}$ and $10^{13}$ cm$^{-2}$. The charge traps at the interface oxide / etched silicon may be gradually filled according to the surface potential, which depends both on $V_{gs}$ and $V_{ds}$. This additional charge induces a modification of the flat band voltage and the vertical electrical field leading to both threshold voltage variation and slop variation. In the Figure 3 b), we observe a quite large dispersion of raw data which makes the comparison difficult. Anyway, the computed values are in the good order of magnitude. The maximum of transconductance, is around 1 µS for ($V_g$=6.5V, $V_{ds}$=2.75V) leading to an $I_d$ current of about 3 µA. For this operating point, $\mu_{eff}$ was evaluated at 300 cm$^2$/V.s.

Let us consider a device on which dynamic resonance were observed. The measured characteristics are $L_G$=10 µm, w=165 nm, g=120 nm, $L_{MOS}$= 6.8 µm, $W_{MOS}$=400 nm, $t_{ox}$=2 nm, and $N_a$= 5.10$^{15}$ at/cm$^3$ [12]. RF characterizations were performed using a measurement bench with a vectorial network analyser Agilent 8753E (VNA) to measure the $S_{12}$ parameter (ratio of the transmitted power over the incident power in the logarithmic scale). The values of the bias voltages applied on the component were extracted from the static characteristics $I_d(V_g)$ and $I_d(V_{ds})$. Electrode voltage $V_{dc}$, gate voltage $V_g$ and drain voltage $V_{ds}$ were respectively $V_{dc}$=10V, $V_g$=4.6V, $V_{ds}$=3V. The input power on the electrode was fixed at -41 dBm (ac- voltage around 4.5 mV RMS over 50 Ω). Pressure was fixed at $10^{-6}$ Torr – The quality factor Q was evaluated to ~700 that seems to be a typical value for such a vibrating beam. Our AC-simulation is based on the model presented in precedent section considering the experimental operating point. The total noise floor of the measurement chain loaded with the component is calculated to know the theoretical background level of the $S_{12}$ parameter. This noise results from three main noise sources (uncorrelated noise assumption): the white mechanical noise of the beam, the thermal noise current of the MOSFET and the VNA noise. The mechanical noise is computed using theoretical characteristics of the beam. The electrical noise is computed from the transconductance $g_m$, and the source gate capacitance $C_{gs}$ with our electrical model. $g_m$, $C_{gs}$ and $\mu_{eff}$ are respectively evaluated at 200 nS, 0.2 fF and 100 cm$^2$/V.s. The VNA noise is known through an open loop calibration without device (~380 pA/√Hz). Theoretical results and measurements are shown in the Figure 4. The model is well in agreement with the experiment despite the only use of physical parameters. The theoretical resonance frequency is quite close to the experimental frequency. The difference between the theoretical level and the measured level is only of 1 dB higher. The background shown in the Figure 4 is only due to the VNA. A low distortion of the experimental resonance peak, which is not anticipated by the model, is also visible. Its origin is not yet explained.

We detailed a unified electromechanical model for a LRG-MOSFET only based on physical assumptions. No any extra mathematical parameters to fit the behavior of the device are thus required. For static study, theory seems to be quite well in agreement with the measurement. The visible dispersion might be a poor control at the bulk potential because the body of the MOSFET was kept floating. Other reasons could be a strong surface density of charge traps and a strong roughness of the interface silicon/oxide inducing a large instability of the measurements. Next generation of component is in progress to improve the field effect and the responsivity to the beam movement. Thus, the gap and the channel length are scaled down. A bulk pad is added to avoid any body effect. The gap reduction implies to take into account the Casimir force [13] that may have an impact at tiny gap. This force is already included in the model. Compared with other models, our model is more comprehensive since it computes the charges densities in the channel according to the mechanical oscillations. It may finally be suitable for any electromechanical components based on a field effect transduction.

Authors acknowledge financial support from the French ANR - NANORES contract.

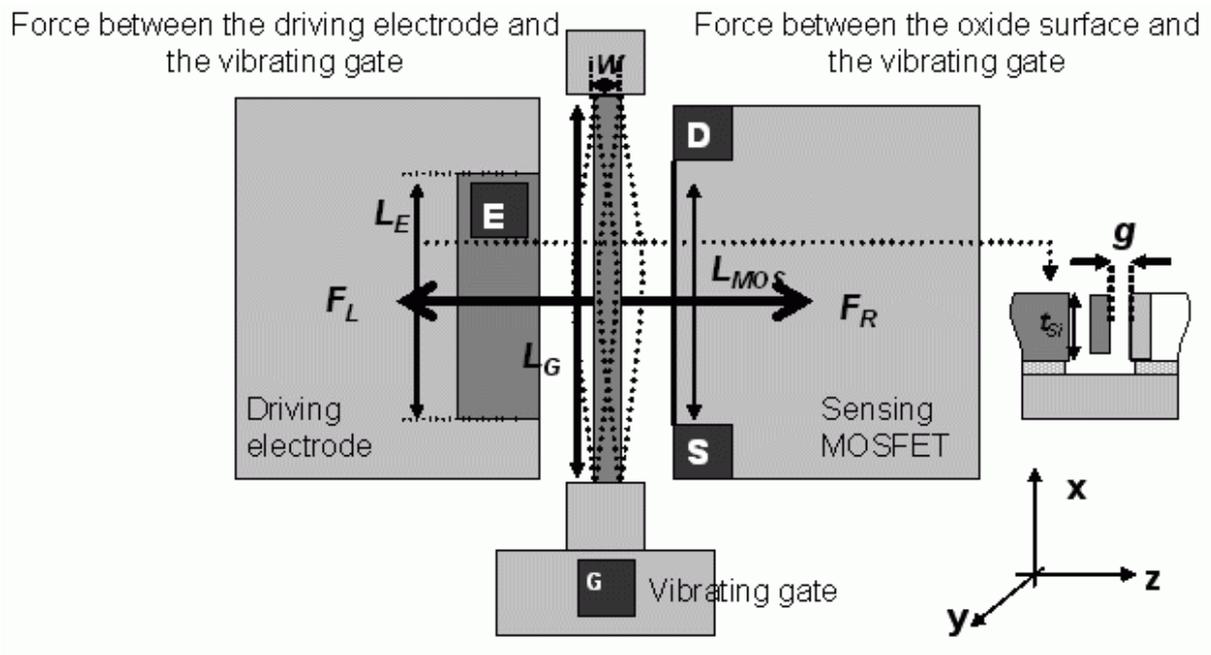

Figure 1: Schematic of the LRG-MOSFET – $F_L$ and $F_R$ are the electrostatic forces on each side of the vibrating gate – $L_E$ and $L_{MOS}$ are respectively the electrode length and the channel length – $g$ is the air gap – $t_{Si}$ is the thickness of the top layer – $w$ is the gate width.

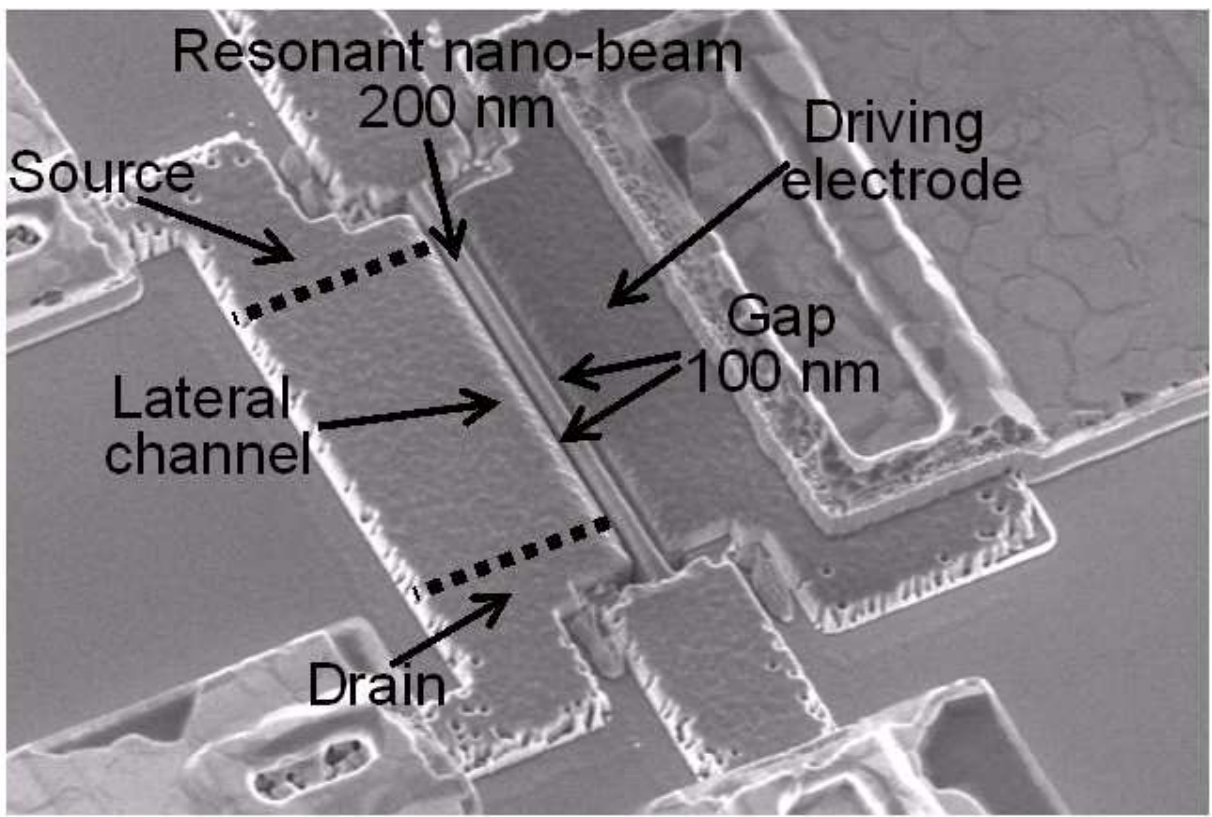

Figure 2: SEM microphotograph of the LRG-MOSFET [Ref 12]

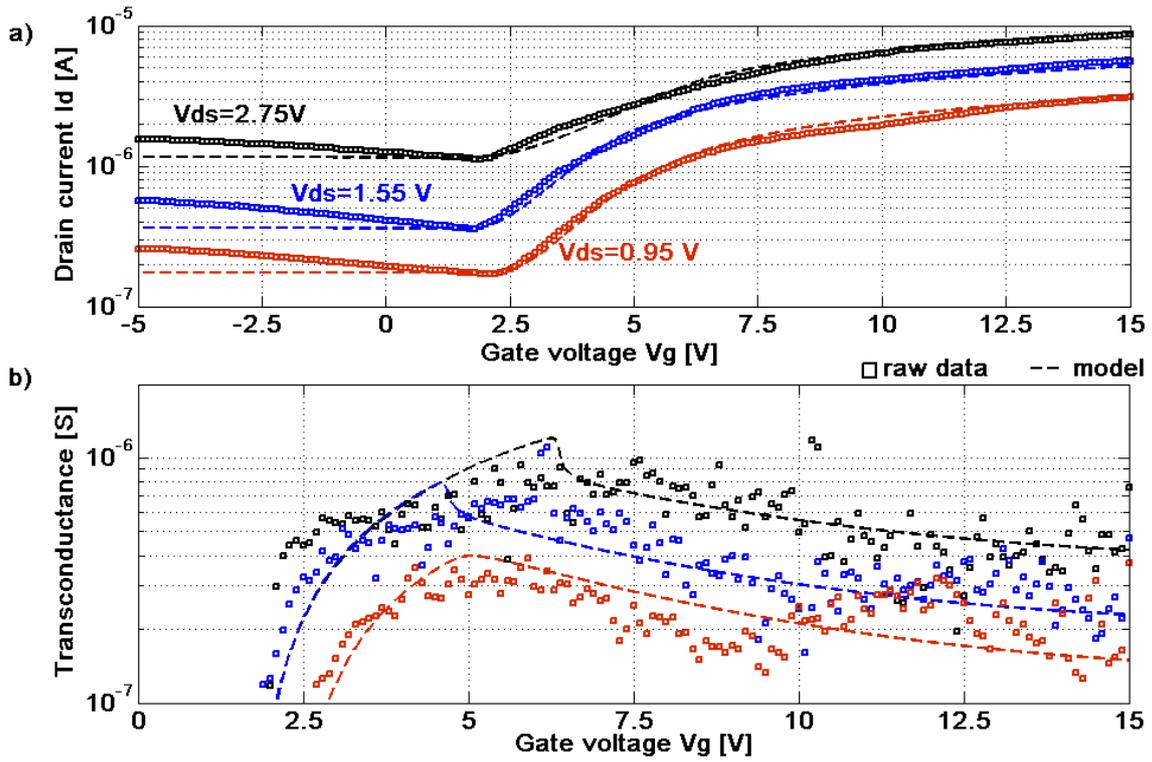

Figure 3: Static characteristics of the LRG-MOSFET – a) *Id(Vg)* – b) Transconduction $g_m(Vg)$.

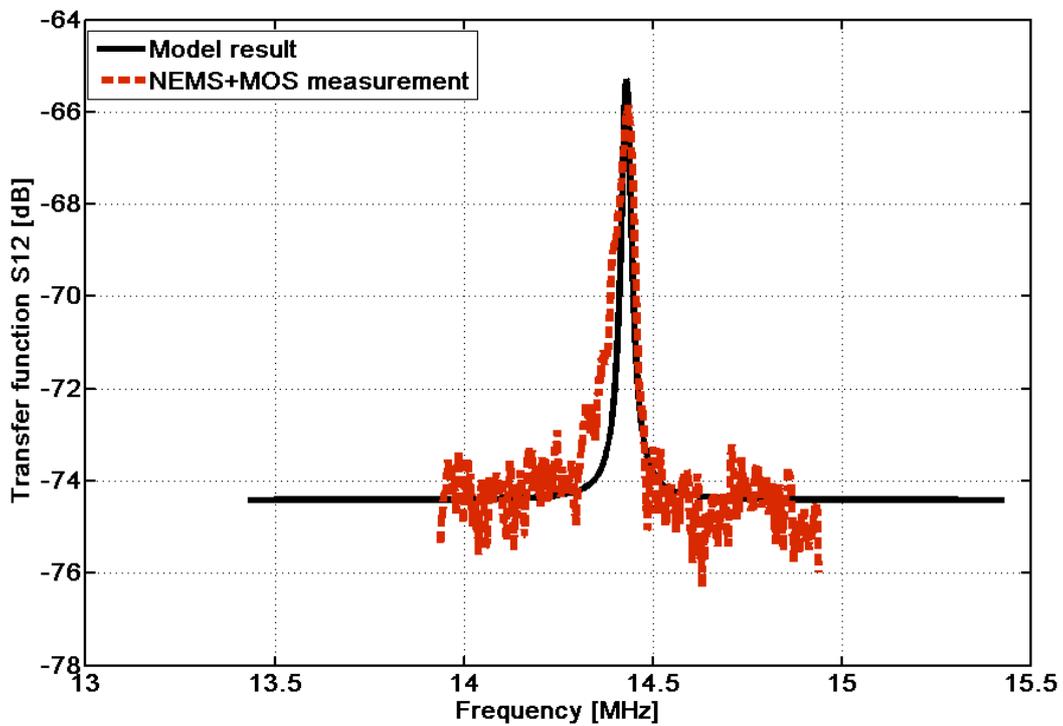

Figure 4: Measured $S_{12}$ parameter and $S_{12}$ computed with our electromechanical model